\documentclass[usenatbib,a4paper]{mn2e}

\pdfoutput=1 

\usepackage{hyperref}
\usepackage{amsfonts}
\usepackage{amssymb}
\usepackage{amsmath}
\usepackage{graphicx}
\usepackage{color}
\usepackage{float}
\usepackage[totalwidth=505pt,totalheight=680pt]{geometry}

\newcommand{\be}{\begin{equation}}
\newcommand{\ee}{\end{equation}}
\newcommand{\bea}{\begin{eqnarray}}
\newcommand{\eea}{\end{eqnarray}}
\newcommand{\nn}{\nonumber}
\newcommand{\di}{\text{d}}
\newcommand{\bx}{{\bf x}}
\newcommand{\bs}{{\bf s}}
\newcommand{\bz}{{\bf z}}
\newcommand{\bv}{{\bf v}}
\newcommand{\bu}{{\bf u}}
\newcommand{\bk}{{\bf k}}

\newcommand{\adsurl}[1]{\href{#1}{ADS}}
\providecommand{\url}[1]{\href{#1}{#1}}

\title[Linear point BAO]
      {Beating non-linearities: improving the
      Baryon Acoustic Oscillations with the linear point}
       
       \author[S.~Anselmi, G.~D.~Starkman, R.~K.~Sheth]
{Stefano Anselmi,$^{1}$\thanks{E-mail: stefano.anselmi@case.edu}
 Glenn D.~Starkman$^{1}$\thanks{E-mail: glenn.starkman@case.edu}
 \& Ravi K.~Sheth$^{2,3}$\thanks{E-mail: shethrk@physics.upenn.edu}\\
 $^1$ Department of Physics/CERCA/Institute for the Science of Origins, Case Western Reserve University, Cleveland, OH 44106-7079 -- USA\\
 $^2$ Center for Particle Cosmology, University of Pennsylvania, 209 S. 33rd St., Philadelphia, PA 19104 -- USA\\
 $^3$ The Abdus Salam International Center for Theoretical Physics, Strada Costiera, 11, Trieste 34151 -- Italy}

\begin{document}
\pagerange{\pageref{firstpage}--\pageref{lastpage}}

\maketitle 

\label{firstpage}

\begin{abstract}
We propose a new way of looking at the Baryon Acoustic Oscillations in the Large Scale Structure clustering correlation function. We identify a scale $s_{LP}$ that has two fundamental features: its position is insensitive to non-linear gravity, redshift space distortions, and scale-dependent bias at the $0.5\%$ level; it is geometrical, i.e.~independent of the power spectrum of the primordial density fluctuation parameters. These two properties together make $s_{LP}$, called the ``linear point'', an excellent cosmological standard ruler. The linear point is also appealing because it is easily identified irrespectively of how non-linearities distort the correlation function. Finally, the correlation function amplitude at $s_{LP}$ is similarly insensitive to non-linear corrections to within a few percent.  Hence, exploiting the particular Baryon features in the correlation function, we propose three new estimators for growth measurements. A preliminary analysis of $s_{LP}$ in current data is encouraging.
\end{abstract}

\begin{keywords}
large-scale structure of Universe
\end{keywords}

\section{Introduction}
\label{intro}

Soon after the revolutionary discovery, through the careful observation of type Ia supernovae, that the expansion of our universe is accelerating \citep{Perlmutter:1998np,Riess:1998cb}, several other probes were proposed to verify and quantify this important phenomenon. One of the most useful employs the so called Baryon Acoustic Oscillations (BAO). Acoustic waves propagated through the ionized plasm of the early universe, and then were quickly damped at a redshift $z_d\sim 1050$.  In linear theory, the baryon distribution retains some memory of this process \citep{1970ApJ...162..815P}.  This has motivated using the acoustic horizon scale $r_{d}$ as a cosmological standard ruler \citep[e.g.][]{Bassett:2009mm}.  The length scale is imprinted, for example, in the Large Scale Structure (LSS) that we measure today \citep{Eisenstein:2005su}; it produces a peak in the correlation function (CF) of a variety of tracers of the dark matter distribution at $r\sim 100$ Mpc/h \citep[e.g.][and references therein]{2014MNRAS.441...24A,2015AA...574A..59D}. 

Like most standard rulers, the acoustic horizon is imperfect.  Several effects slightly shift or distort the peak in the CF, including linear physics \citep{2007ApJ...665...14S,2008MNRAS.390.1470S}, non-linear gravitational evolution \citep{Crocce:2007dt}, redshift space distortions and scale-dependent bias \citep{2008PhRvD..77d3525S}.  The resulting systematic effects are a serious challenge to using the BAO peak for sub-percent precision cosmology.  Many approaches have been suggested to mitigate these effects.  \citep[A non-exhaustive list includes][.]{Crocce:2007dt, Taruya:2007xy, Anselmi:2012cn, 2010PhRvD..82j3529D, 2015arXiv150404366B}

In this paper we propose a new way of looking at the BAO features in the correlation function. We identify a point in the correlation function that has three remarkable properties:
\begin{itemize}
\item[(1)] it is geometric, i.e.~its position is independent of the power spectrum of the primordial density fluctuation parameters;
\vspace{0.2cm}
\item[(2)] it is unaffected by nonlinearities and scale-dependent bias at percent-level precision, i.e.~its position is redshift-independent;
\vspace{0.2cm}
\item[(3)] the amplitude of the matter and halo correlation functions at this point are very weakly affected by nonlinearities or scale-dependent bias.
\end{itemize}
This point lies midway between the scales associated with the peak and the dip in the correlation function on BAO scales, a definition that holds true even as the dip and the peak shift due to nonlinear physics. 
Given (2) and (3) we denote this the ``linear point'' (LP).
Properties (1) and (2) make the LP a standard ruler;
property (3) makes the LP a useful window into growth of structure and bias.
We discuss both below.

We devote most of this paper to an investigation of the nonlinear corrections to all the three relevant BAO scales, i.e.~the dip, the peak and the linear point, taking into account the dominant effects of nonlinear gravity, redshift-space distortions and scale-dependent bias. The three scales show  different sensitivities to these effects.  We find the peak to be the most shifted, while the dip is shifted in the opposite direction.  Due to the near-antisymmetry of the CF across the linear point, the position of the LP is much more stable.  This makes the LP a potentially useful standard ruler.  Furthermore, the matter and halo correlation functions evaluated at the LP agree with the linear theory values to a few percent. We exploit these features to propose new ways of measuring the linear growth and linear bias, and to provide consistency checks of our findings with present and future data. 

This paper is organized as follows. In Section \ref{sec:lin} we introduce the peak, the dip and the ``linear point'' in the linear theory correlation function of dark matter.  We show that the linear point is almost a purely geometrical point, emphasizing how the correlation function is nearly antisymmetric about this linear point. In Section \ref{sec:nl} we introduce the main sources of nonlinear distortions that affect the three BAO scales, namely nonlinear gravity, redshift distortions and scale dependent bias. We quantify systematics from nonlinear evolution in Section \ref{sec:BAO} by showing how the near-antisymmetry of the CF across the linear point is the key feature that drives the linear behavior of the linear point. The correlation function amplitude is also linear to a few percent at the dip-peak middle point. Section \ref{sec:use} discusses a number of possible estimators of the linear point.  It also leverages the linear theory behaviour of the LP to furnish estimates of the linear theory growth factor.  A final section summarizes our findings and discusses future directions.

\section{The geometric ``linear point'' in the correlation function}
\label{sec:lin} 

There were two principal ideas behind Baryon Acoustic Oscillation (BAO) measurements: 
\begin{description}
  \item[(i)] the comoving baryon acoustic scale at the Compton drag epoch $r_{d}$ corresponds to the baryon acoustic peak $s_{p}$ of the matter correlation function;
  \vspace{0.3cm}
  \item[(ii)]  Baryon Acoustic Oscillation measurements are a purely geometrical probe, because the baryon acoustic scale contains no information about  primordial density fluctuations, and so is independent of the initial amplitude $A_{s}$, the spectral index $n_{s}$ and the optical depth to the last scattering surface $\tau$.
\end{description}  
It is well known that both these assumptions break down at the linear \citep{2007ApJ...665...14S,2008MNRAS.390.1470S} and nonlinear levels \citep{Crocce:2007dt,2008PhRvD..77d3525S} when one probes the BAO at a precision of a few percent, which is relevant in this age of precision cosmology.

While it is well known that $s_p$ does not quite equal $r_d$, in the following we will identify an actual linear geometric point in the correlation function whose position is not affected by nonlinear distortions at the percent level:  it does not shift with time, i.e.~redshift. We name this point the ``linear point'' (LP).  The linear point is therefore an actual standard ruler, unlike the peak scale $s_p$ in the correlation function.  

\begin{figure}
\centering
\includegraphics[width=1\hsize]{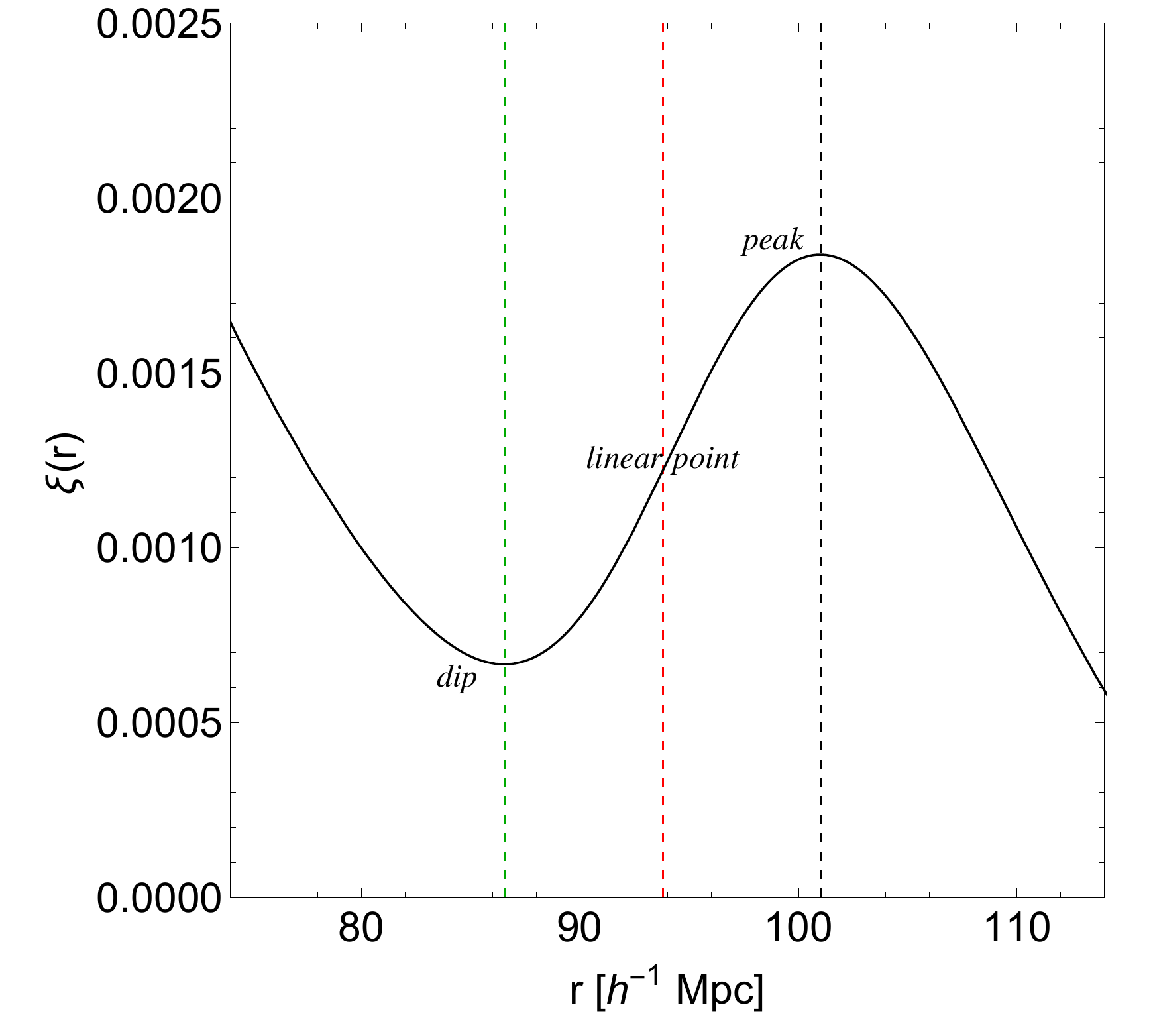}
\caption{\label{fig:BAOfeature} BAO scales in the correlation function:  the peak, a dip and the mid-point between the two, which we call the ``linear point'' (see text).}
\end{figure}

\begin{figure*}
\centering
\includegraphics[width=1.0\hsize]{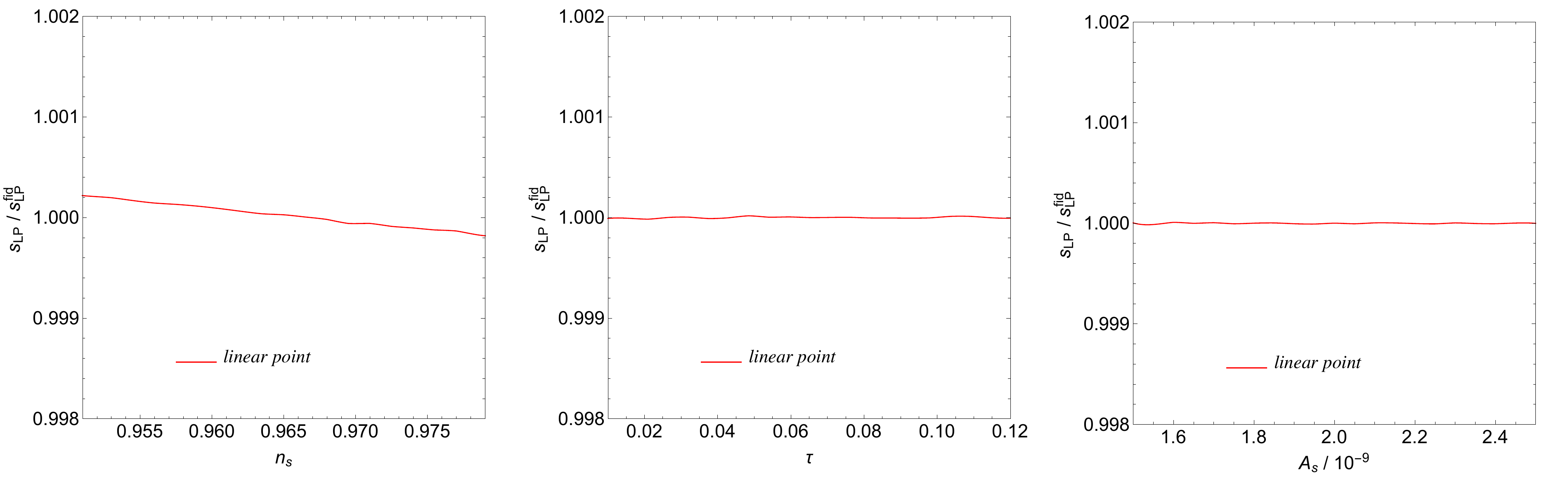}
\caption{\label{fig:prim:fluct} Dependence of the ``linear point'' $s_{LP}$ on the primordial density fluctuation parameters, $n_{s}$, $\tau$ and $A_{s}$, expressed relative to its value in a fiducial model, which we take to be that of the \cite{2015arXiv150201589P}.  Smooth red line shows that $s_{LP}$ is independent of  $n_{s}$, $\tau$ and $A_{s}$ to the $0.02\%$ level.}
\end{figure*}

We start our analysis by introducing in Fig.~\ref{fig:BAOfeature} three scales in the BAO correlation function:  a peak, a dip and the ``linear point,'' defined (for reasons we explain below) as the peak-dip middle point:  $s_{LP}\equiv (s_d + d_p)/2$.  The scale $s_{LP}$ corresponds to the inflection point in $\xi$ to better than $0.5\%$ precision; this will be important when we consider scale-dependent bias in the next sections.  \footnote{In what follows, we focus on an analysis of $\xi$.  In principle, one might have thought that $r^2\xi$ was at least as appropriate.  We discuss shortly why we have chosen $\xi$ instead.}  

In Fig.~\ref{fig:prim:fluct}, by performing a linear analysis we show that the position of the linear point is indeed independent of $n_{s}$, $\tau$ and $A_{s}$  at the $0.02\%$ level. (The $A_{s}$ independence is true by definition in linear theory. Furthermore we checked that varying simultaneously $n_{s}$ and $\tau$ does not change the results presented in Fig.~\ref{fig:prim:fluct}.)  \footnote{We set the cosmological parameters $\Omega_{b}=0.0486$, $\Omega_{c}=0.259$ and $H_{0}=67.74$ to be those of the fiducial model from the \cite{2015arXiv150201589P}, and we compute the linear matter power spectrum using the CAMB code \citep{2000ApJ...538..473L}.  While the scales $s_d$, $s_p$ and $s_{LP}$ depend on these parameters, the `geometrical' nature of $s_{LP}$ is generic.}
While this geometric property holds true in linear theory for the dip and the peak scales as well, it is well known that nonlinear evolution affects the positions of the peak and the dip, spoiling their geometric nature and their redshift independence \citep{Crocce:2007dt,2008PhRvD..77d3525S}, and thus their status as standard rulers. 

The linear correlation function is anti-symmetric in the peak-to-dip range (see Fig. \ref{fig:BAOfeature}).  One measure of this symmetry is to compare the scale $s_{LP}$ which was defined by the peak-dip positions $(s_d + d_p)/2$, with the scale $s_A$ on which the amplitude of $\xi^{lin}$ equals $[\xi^{lin}(s_{d})+\xi^{lin}(s_{p})]/2$.  These two scales differ only slightly:  $s_A\approx 1.002s_{LP}$.   In addition, the amplitude $\xi^{lin}(s_{LP})$ corresponds to $\xi^{lin}(s_A)$ to $\sim 2-3\%$ precision.  (We show below that nonlinear effects improve this amplitude agreement to $1\%$ level.)
This anti-symmetry will be crucial for understanding why the linear point $s_{LP}$ is nearly unaffected by nonlinearities, and is the main reason why we prefer to work with $\xi$ rather than $r^2\xi$.

The linear-theory analysis presented in this section thus reveals that the matter correlation function gives us a new geometric point in the correlation function, the dip-peak midpoint, called the ``linear point.'' In the next sections, we show that the linear point is unaffected by nonlinearities at percent level precision, making it an excellent cosmological ``standard ruler.''

\section{Nonlinear distortions}
\label{sec:nl}

\subsection{Real space}

Nonlinear gravitational clustering smooths the peak in the correlation function and shifts it slightly \citep{1996ApJ...472....1B,2007ApJ...665...14S,Crocce:2007dt}.  The smoothing is due to displacements from the initial positions, so the one-dimensional velocity dispersion in linear theory, 
\be
	\sigma_{v}^{2}(z)=\frac{1}{3} \int \frac{\di^{3} q}{(2 \pi)^{3}}\frac{P^{lin}(q,z)}{q^{2}} \, , 
	\label{sigma:v}
\ee
where $P^{lin}$ is the matter power spectrum (PS) in linear theory, sets the scale of the smoothing.  

In particular, in the Zel'dovich approximation \citep{Zeldovich:1969sb}, the smoothing due to the displacements is well-approximated by 
\be 
	P^{nl}(k) \approx e^{- k^{2} \sigma_{v}^{2}} P^{lin}(k)\, 
	\label{PSzel}
\ee
\citep{Crocce:2005xy, 2015arXiv150507477P, 2015arXiv150605264V}.  
The resulting equation for the nonlinear $\xi$ is 
\be
	\xi^{nl}(r) \approx \int \frac{\di k}{k}\,
      \frac{k^{3} P^{lin}(k)}{2 \pi^{2}}\,e^{- k^{2} \sigma_{v}^{2}(z)}\, j_{0}(kr)\, .
	\label{xiZel}
\ee
Equation~(\ref{xiZel}) has been shown to accurately describe the correlation function in the BAO regime \citep{Crocce:2007dt, Angulo:2007fw, 2014MNRAS.440.1420R, 2015arXiv150605264V, 2015arXiv150507477P, 2015arXiv150404366B} and it has successfully been employed  for BAO reconstruction \citep{2007ApJ...664..660E, 2015MNRAS.450.3822W}.

\subsection{Redshift space}
Galaxies are observed in redshift space, where the Hubble flow induced cosmological redshift has been distorted by peculiar velocities.  The real-space-redshift-space map in the plane-parallel approximation is 
\be
	\bs = \bx - fv_{z}\hat{\bz}\,.
	\label{srmap}
\ee
Here $f=\di \ln D/\di \ln a$ is the growth rate, 
$\hat{\bz}$ is the line of sight direction, 
and $\bv(\bx)=-\bu/(\mathcal{H}f)$, 
where $\bu(\bx)$ is the peculiar velocity and $\mathcal{H}$ is the conformal Hubble parameter.   

The same bulk motions which generate nonlinear clustering also give rise to redshift space distortions \citep{Kaiser:1987qv}.  Moreover, the angular dependence of equation~(\ref{srmap}) means that the clustering signal can (and should!) be decomposed in spherical harmonics.  For example, the redshift-space monopole associated with $\xi^{nl}$ is 
\bea
	\xi^{s,\, nl}_{0}(s) = \frac{1}{2} \int_{-1}^{1} \di \mu 
  \int \frac{\di k}{k}\, \frac{k^{3}P^{lin}(k)}{2 \pi^{2}} (1+\mu^{2}f)^{2}\,\nn \\
	  \qquad \times\quad \,e^{- k^{2} \sigma_{v}^{2}(1+\mu^{2}f(2+f))} j_{0}(ks) \, ,
	\label{red:nl}
\eea
where $\mu=\hat{\bk}\cdot\hat{\bz}$ is the cosine between the line of sight and the wave vector.  The integral over $\mu$ can be performed analytically, making numerical integration over $k$ very fast \citep{2010PhRvD..81b3526D, 2015arXiv150507477P}.  The linear theory limit has no nonlinear smearing; so, setting $\sigma_{v}\to 0$ yields $\xi^{s}_{0}(s) = (1+ 2f/3+{f^{2}/5})\,\xi(s)$ in linear theory \citep{Kaiser:1987qv}.

\subsection{Halo bias: linear and scale dependent}

In the real Universe (as opposed to simulations thereof), we cannot directly observe the matter field.  Instead, we measure galaxies that inhabit halos which, in turn, are biased tracers of the matter field. In this section we focus on how halo bias (the halo-matter relation) affects the correlation function in the BAO regime.  A detailed analysis of how bias affects the BAO features of the correlation function is beyond the scope of this paper. However, we present a preliminary investigation of this topic, and expect to carry on this study in future works.

We use the peaks theory approach to halo bias \citep{1986ApJ...304...15B}.  As before, we employ a simplified description that allows us to focus on the main effects (e.g., we ignore mode coupling terms).  A good approximation to the correlation function of peaks is 
\bea
	\xi^{nl}_{hh}(r) = \int \frac{\di k}{k} \frac{k^{3}P^{lin}(k)}{2 \pi^{2}}\,e^{- k^{2} \sigma_{v}^{2}} \qquad\qquad\nn \\
	\qquad\qquad\times\ \,\left[b_{10}^{E}(z)+b_{01}^{E}(z)k^{2}\right]^{2}\, j_{0}(kr), \qquad\qquad
	\label{peak:nl}
\eea
where $b_{10}^{E}$ is the usual Eulerian linear bias \citep{2010PhRvD..82j3529D}.  Similarly, the redshift-space monopole is simply
\bea
	\xi^{s,nl}_{0, hh}(s) = \frac{1}{2} \int_{-1}^{1} \di \mu \int \frac{\di k}{k} \frac{k^{3}P^{lin}(k)}{2 \pi^{2}} \qquad\qquad\qquad\qquad\qquad \nn \\
	\quad\times \left[b_{10}^{E}(z)+b_{01}^{E}(z)k^{2}+\mu^{2} f\right]^{2}	 \, e^{- k^{2} \sigma_{v}^{2}(1+\mu^{2}f(2+f))}\, j_{0}(ks)
	\label{peak:nl:s}
\eea
where the integral over $\mu$ can again be done analytically \citep{2010PhRvD..81b3526D}.  

Strictly speaking, we should replace $\sigma_v$ in the expressions above with the velocity-biased value appropriate for peaks:  $\sigma_p<\sigma_v$ \citep{2010PhRvD..81b3526D}.  In addition, the expressions above should include the smoothing windows of radius $R_{p}$ (one for each peak) associated with identifying the peaks.  However, these are almost degenerate with $\sigma_p$ in real space (and not quite as denegerate in redshift space).  In effect, the expressions above have set $R_p^2+\sigma_p^2 \approx \sigma_v^2$.  In this respect, we compute $\sigma_{v}$ from equation~(\ref{sigma:v}) while we treat equations~(\ref{peak:nl}) and~(\ref{peak:nl:s}) as effective formulae, where $b_{10}^{E}$, $b_{01}^{E}$ could be evaluated by fitting to simulations \citep[as was done in][]{2015arXiv150507477P}.

\section{Why does the Linear Point remain linear?}
\label{sec:BAO}

\subsection{Dark matter}
\label{subsec:m}
Nonlinearities affect the locations of the BAO peak, dip and LP scales in two different ways:
\begin{itemize}
 \item[(a)] smoothing the correlation function;
 \item[(b)] shifting the correlation function
\end{itemize}
\citep[e.g.][]{2008PhRvD..77d3525S}.  To see this clearly, rewrite equation~(\ref{xiZel}) as 
\be
	\xi^{nl}(r) = \xi^{nl}(\vert \bx \vert;R_G)\quad\,\label{xi:conv}
\ee
where $R_{G}= \sqrt{2}\, \sigma_{v}$ and 
\be
	\xi^{nl}(\vert \bx \vert;R) = \int \di^{3} x' \, 
          \frac{e^{-\frac{\vert \bx-\bx' \vert^{2}}{2 R^{2}}}}{(2 \pi R^{2})^{3/2}}\, 
          \xi^{lin}(\vert \bx' \vert).
\ee
The 3D convolution can be simplified to
\bea
        \xi^{nl}(\vert \bx \vert;R) =
	  \int \di r' \, \frac{r'}{r}\frac{e^{-\frac{(r-r')^{2}}{2 R^{2}}}-e^{-\frac{(r+r')^{2}}{2 R^{2}}}}{(2 \pi R^{2})^{1/2}}\, \xi^{lin}(r')  \nn \\ 
	  \backsimeq \int \di r' \, \frac{r'}{r}\frac{e^{-\frac{(r-r')^{2}}{2 R^{2}}}}{(2 \pi R^{2})^{1/2}}\,\xi^{lin}(r')\,,  \qquad\qquad
	\label{conv1D}
\eea
where the last equality is practically exact for $r>2R$. 
This shows explicitly that the exponential term inside the integral smooths the correlation function, while the $r'/r$ term shifts the peak of the Gaussian filter to higher $r$.  This, in turn, moves the peak of $\xi$ to smaller $r$.  

Similarly, equation~(\ref{red:nl}) for the redshift-space monopole can be written as 
\be
	\xi_{0}^{s,nl}(s) = \frac{1}{2} \int_{-1}^{1} \di \mu\, (1+\mu^{2}f)^{2}\,
        \xi^{nl}(\vert \bx \vert;S_{G})\label{xis:conv}
\ee
where $S_{G}=\sqrt{2} \,\sigma_{v} \, \sqrt{1+\mu^{2}f(2+f)}$.  
Since $S_{G}>R_G$ the smoothing is larger in redshift space, but the overall effect is otherwise similar to that in real space.

\begin{figure}
\centering
\includegraphics[width=1.0\hsize]{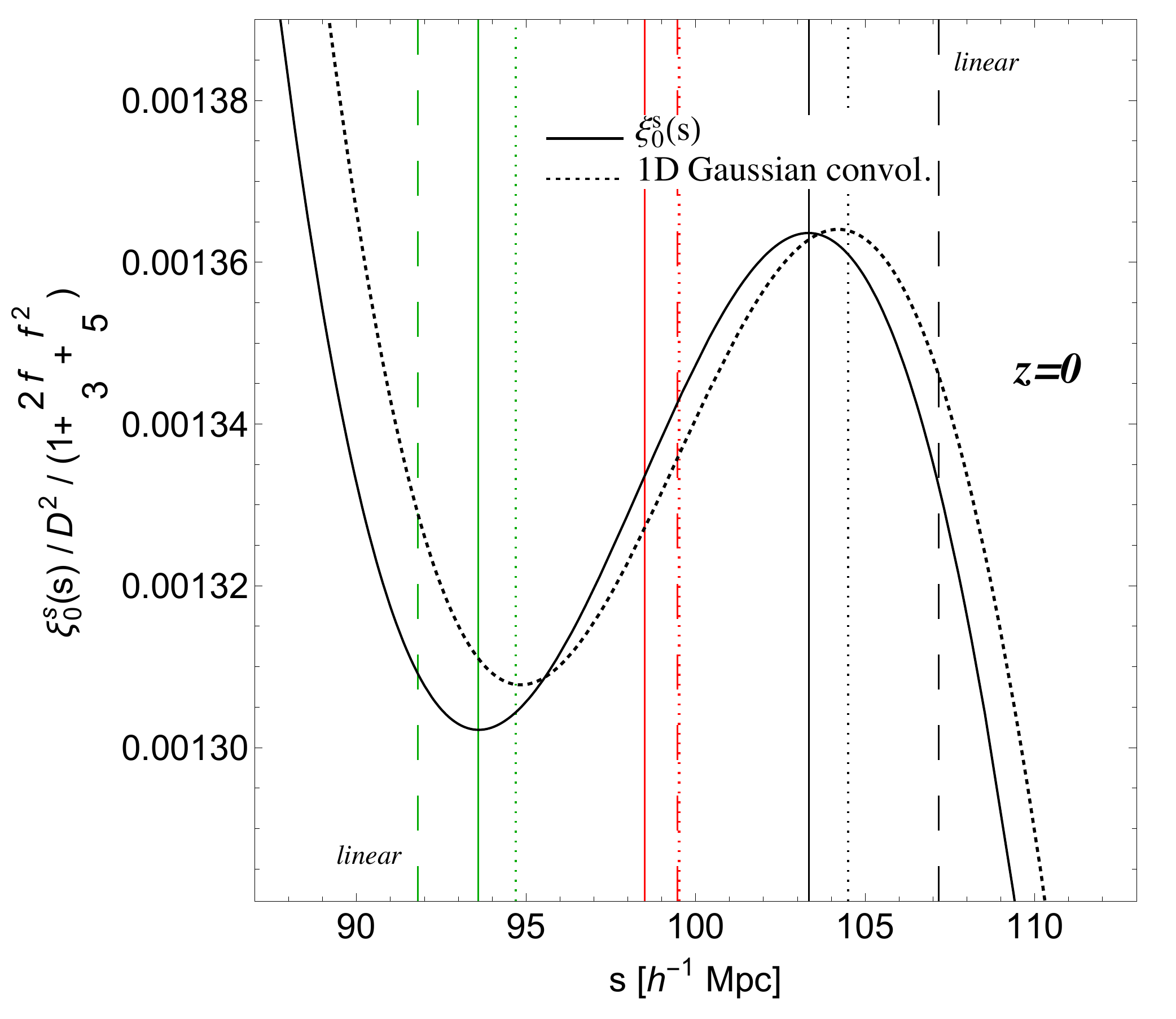}
\caption{\label{fig:shift} Monopole of the redshift-space correlation function rescaled by the linear Kaiser factor of $(1 + 2f/3+ f^{2}/5)$. Vertical lines show the scales of the peak, dip and linear point for linear theory (dashed), the 1D Gaussian convolution (dotted) (i.e.~equation~(\ref{conv1D}) with $r'/r \to 1$), and the nonlinear model $\xi_{0}^{nl}(r)$ (continuous).}
\end{figure}

To illustrate the effect of these nonlinearities, Fig. \ref{fig:shift} shows the monopole of the redshift-space correlation function rescaled by the linear theory factor of $D^2\,(1 + 2f/3 + f^{2}/5)$ at $z=0$ in a $\Lambda$CDM cosmology with $\Omega_{m}=0.265$, $\Omega_{b}h^2=0.0226$, $h=0.71$, $n=0.963$ and $\sigma_{8}=0.8$.\footnote{Note that these are not the Planck cosmological parameters we used in Section~\ref{sec:lin}; rather, they are those of the simulation set we use later in the paper (see Section \ref{sec:sim}). This is why, the peak, dip and LP scales are $\sim 5h^{-1}$Mpc larger than in Fig.~\ref{fig:BAOfeature}.}  The vertical lines show the positions of the peak,  dip and linear point for linear theory ($\sigma_v\to 0$) (dashed), the 1D Gaussian convolution (i.e.~equation~(\ref{conv1D}) with $r'/r \to 1$) (dotted), and the nonlinear model $\xi_{0}^{nl}(r)$ (continuous). 
Notice that the 1D Gaussian convolution moves the peak and the dip by the same amount, but in opposite directions, whereas the $r'/r$ term shifts the whole correlation function to smaller values of $r$ by $1\%$.  Thus, the two effects add for the peak but they subtract for the dip.  As a result, the dip shifts less than the peak.  

More importantly, the scale of the linear point shifts even less than either the peak or the dip.  The net result is that the peak moves by $3.5\%$ to the left, the dip by $1.5\%$ to the right and the linear point by just $1\%$ to the left.  (These shifts are smaller at higher redshifts, of course.)  The peak is consequently the most sensitive point to nonlinear corrections and the linear point is the least.  Moreover, because the dip and the peak share the same nonlinear smoothing, the amplitude of the linear point remains unchanged to within a few percent even though the dip and the peak (heights and scales) change significantly.  
\subsection{Biased tracers:  Halos}
To understand how halo bias affects the correlation function in the BAO regime we focus, for simplicity, on the real space results. The same arguments apply to the redshift space monopole. 

To gain insight we write equation~(\ref{peak:nl}) as a convolution of the linear correlation function, and simplify as before to obtain 
\bea
	\xi^{nl}_{hh}(r) 
  &\backsimeq& \int \di r' \, \frac{r'}{r}\ \, \xi^{lin}(r') \,
               \frac{e^{-\frac{(r-r')^{2}}{2 R_{G}^{2}}}}{(2 \pi R_{G}^{2})^{1/2}} \nn \\ 
	&&\times \left\{b_{10}^{2}+2 b_{10}b_{01}\frac{R^{2}_{G}-(r-r')^{2}}{R^{4}_{G}}\right. \nn \\ 
	&&\quad +\left.b_{01}^{2}\frac{(r-r')^{4}-6(r-r')^{2}R^{2}_{G}+3 R^{4}_{G}}{R^{8}_{G}}\right\}\,, 
	\label{halo:conv1D}
\eea
where we have dropped the ``E'' on the bias factors for convenience.

\cite{2010PhRvD..82j3529D} have emphasized the fact that the terms proportional to $b_{01}$ and $b_{01}^{2}$ contribute significantly to the shape of $\xi_{hh}$ on BAO scales.  As their Figure~3 shows, these terms are approximately anti-symmetric in $r$, so they serve to enhance the BAO feature by increasing the height of the peak and depressing the dip.  However, their figure also shows that these terms contribute little to the scale or the amplitude of the correlation function at the linear point -- a point they do not discuss.  

There are two ways to see why this happens.  First, note that the term $\propto b_{01}$ scales as $k^2$ in Fourier space, so it is $\propto \nabla^2\xi^{lin} = (2/r)\,\partial\xi^{lin}/\partial r + \partial^2\xi^{lin}/\partial r^2$ in real space.  On the peak and dip scales, the first term is zero, but the second term contributes maximally to the amplitude of $\xi_{hh}$.  However, at the inflection point, the second term is zero by definition, and the first term is almost surely very small on BAO scales because biased tracers assemble their mass from regions which are much smaller than the BAO scale.  
Hence, at the inflection point, the net contribution from the $b_{01}$ term is negligible.  The argument for the $b_{01}^2$ term is similar:  It scales as $k^4$ in Fourier space, so it is $\propto \nabla^2(\nabla^2\xi) = 4\xi^{(3)}/r + \xi^{(4)}$ (i.e., to third and fourth derivatives of $\xi^{lin}$ wrt $r$).  Since $\xi^{lin}$ is rather smooth at the inflection point, these higher derivatives are small, and so the net contribution from the term $b_{01}^2$ term is small.  Figure~3 in \cite{2010PhRvD..82j3529D} shows that, indeed, while the terms $\propto b_{01}$ and $b_{01}^2$ contribute maximally around the BAO peak and dip scales, they are substantially smaller at the inflection point.  Since $s_{LP}$ lies very close to the inflection point, we expect the scale-dependent bias terms to be strongly suppressed at $s_{LP}$, making $\xi_{hh}^{nl}(s_{LP})\approx b_{10}^2\, \xi^{nl}(s_{LP})$.\footnote{There is another inflection point just beyond the BAO peak-scale.  Therefore, we expect the scale-dependent bias terms to be suppressed there as well, and this is indeed seen in Figure~3 of \cite{2010PhRvD..82j3529D}.  While exploiting this larger scale as an additional standard ruler is beyond the scope of this work, we return briefly to this scale in the next section.}

Alternatively, we can modify the convolution argument of the previous subsection to see how the terms related to $b_{01}$ and $b_{01}^2$ contribute to $\xi_{hh}^{nl}$.  To begin, we note that 
\be
	\int \di r'\,\frac{e^{-\frac{(r-r')^{2}}{2 R_{G}^{2}}}}{(2 \pi R_{G}^{2})^{1/2}}\,
              \frac{R^{2}_{G}-(r-r')^{2}}{R^{4}_{G}} = 0
\ee
and 
\be
	\int \di r'\,\frac{e^{-\frac{(r-r')^{2}}{2 R_{G}^{2}}}}{(2 \pi R_{G}^{2})^{1/2}}\,
              \frac{(r-r')^{4}-6(r-r')^{2}R^{2}_{G}+3 R^{4}_{G}}{R^{8}_{G}} = 0
\ee
also.  
If $\xi^{lin}(r)$ is approximately antisymmetric in $r$, then the integrals above will remain very small even if we insert a factor of $\xi^{lin}(r')$ in the integrands.  The terms proportional to $b_{01}$ and $b_{01}^2$ include an additional $r'/r$ factor, but this factor will not change the value of the integral dramatically.  We have checked numerically that including it indeed makes little difference.  As a result, the terms proportional to $b_{01}$ and $b_{01}^2$ contribute little to the result, in agreement with our discussion in the preceding paragraph \cite[and with Figure~3 of][]{2010PhRvD..82j3529D}.  Thus, this provides another way to see why $\xi_{hh}^{nl}(s_{LP})$ is insensitive to scale-dependent bias even when $b_{01}$ is not negligible, despite the fact that the height and scale of the peak and dip can be strongly modified.  

We conclude that the antisymmetric structure of the correlation function is preserved even in the presence of scale-dependent bias.  While this scale-dependence moves the peak and the dip in opposite directions, the linear point remains fixed.  Furthermore the amplitude of the biased correlation function at the linear point is only sensitive to the scale-independent part of the linear bias factor, $b_{10}$.  Both these reasons increase the appeal of the LP as a standard ruler candidate.

\subsection{Comparison with simulations}
\label{sec:sim}
To check the accuracy of our analysis, we first compare equations~(\ref{xiZel}) and (\ref{red:nl}) to measurements in the N-body simulations of a $\Lambda$CDM cosmology with  $\Omega_{m}=0.265$, $\Omega_{b}h^2=0.0226$, $h=0.71$, $n=0.963$ and $\sigma_{8}=0.8$ carried out by \cite{2011PhRvD..84d3501S}.   

\begin{figure}
\centering
\includegraphics[width=1.0\hsize]{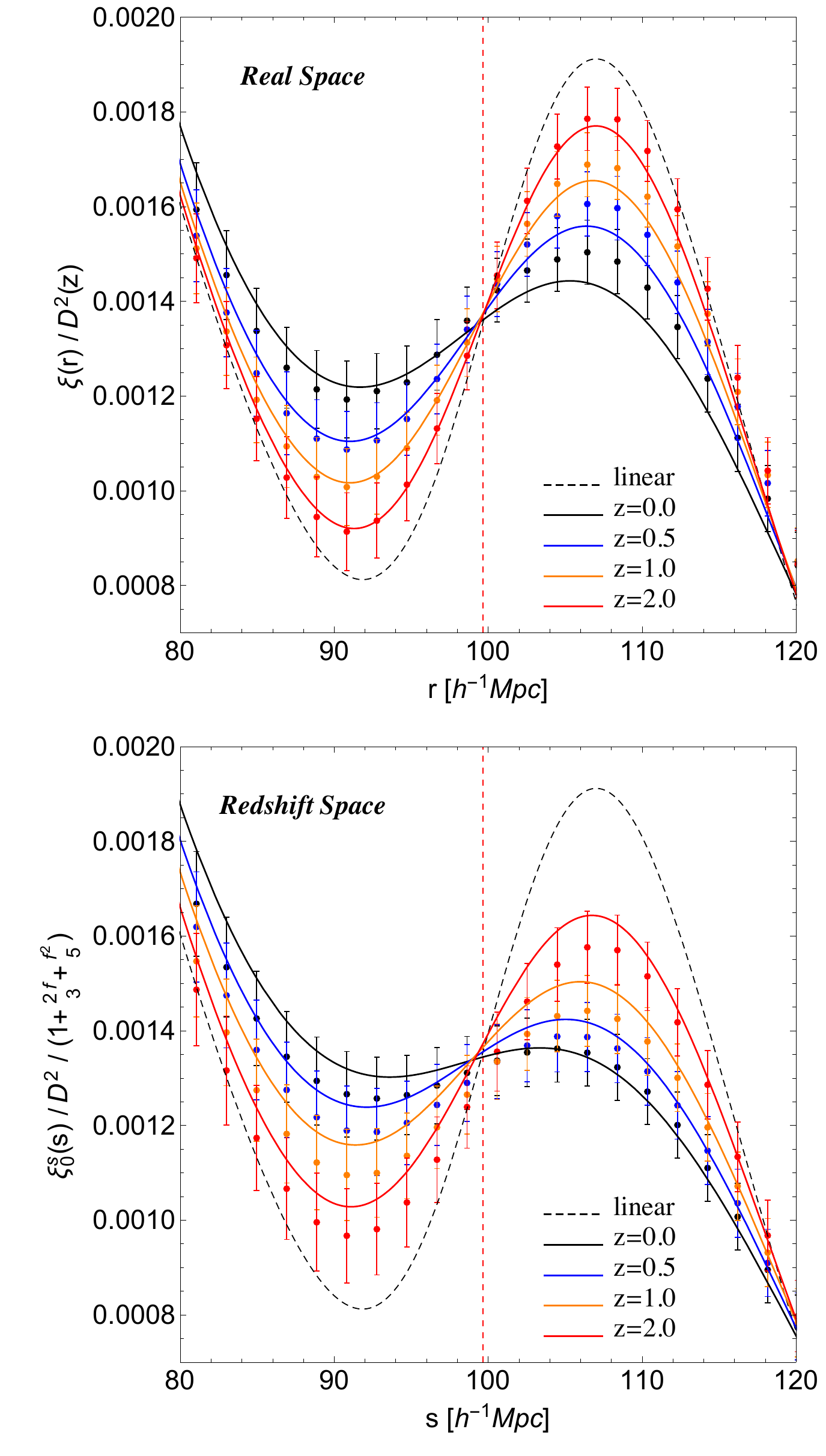}
\caption{\label{fig:rescaled} Top: Real space correlation function rescaled by the linear growth factor at four redshifts.  Bottom:  Similarly rescaled redshift-space monopole.  Smooth lines show equations~(\ref{xiZel}) and (\ref{red:nl}); symbols with error bars show the N-body results. The vertical dashed red line marks the scale of the linear point in linear theory. }
\end{figure}

The top panel of Fig.~\ref{fig:rescaled} shows the real-space correlation function at four different redshifts, rescaled by the scale-independent linear growth factor.  The bottom panel shows the same rescaled quantity for the redshift-space monopole.  The symbols with error bars show the measurements while the continuous lines show the theoretical predictions presented in the previous section.  (We compute these as integrals over $P(k)$ rather than as real-space convolutions.)  

\begin{figure}
\centering
\includegraphics[width=1.0\hsize]{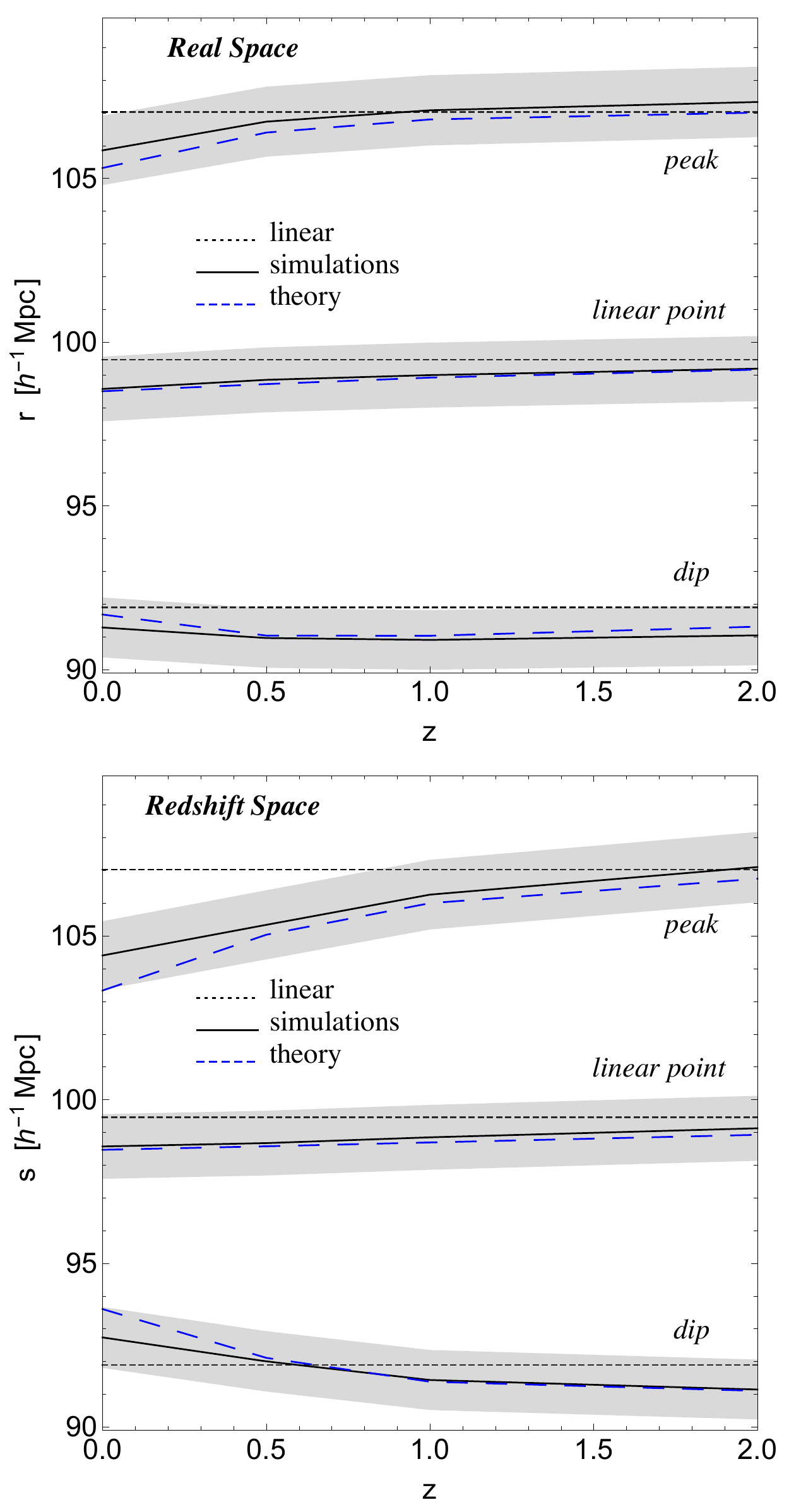}
\caption{\label{sim:theory:m} Top: Real space BAO peak, dip and LP scales. Bottom: Redshift space BAO peak, dip and LP positions.  Smooth black curves show the measurements in simulations, shaded gray regions show $1\%$ variation around the measurements, and dotted and dashed curves show the linear and nonlinear theory values of these scales.}
\end{figure}

The simulation measurements scatter much less than the $5-10\%$ one would expect given the error bars, which were obtained from the scatter among 30 different realizations.  Apparently, the points are strongly correlated, and a full error analysis must properly treat the covariance.  If the smooth curve that the eye would fit through the points is a good approximation of the correlation function to be inferred from the simulations, and the very small point-to-point scatter around that by-eye fit is a fairer approximation of the true error bars, then it appears that, while the amplitude of the correlation function is at times discrepant between the simulations and the theory, the scales of the dip and the peak are correctly predicted to subpercent accuracy, and the LP to $0.5\%$. Fig.~\ref{sim:theory:m} shows this explicitly:  the solid curves show the BAO peak, dip and LP scales, which we estimate by fitting a high-order polynomial to the measured $\xi$, and the corresponding dotted and dashed curves show the linear and nonlinear theory values of these scales.  

\begin{figure}
\centering
\includegraphics[width=1.0\hsize]{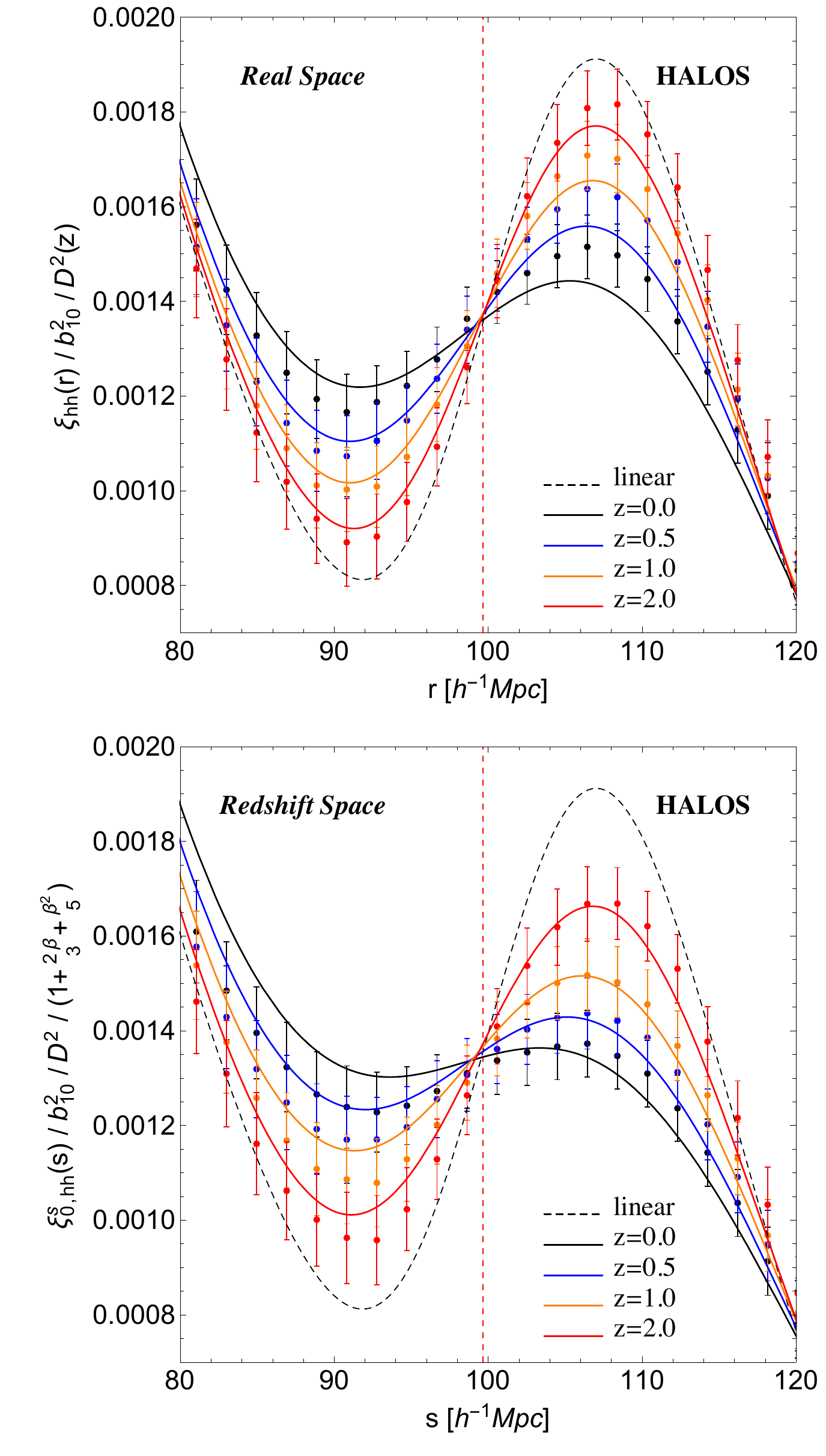}
\caption{\label{fig:rescaled:h} Same as Fig.~\ref{fig:rescaled}, but for biased tracers (halos more massive than $M_{h}>1.37\times 10^{12} h^{-1}M_{\sun}$).}
\end{figure}

\begin{figure}
\centering
\includegraphics[width=1.0\hsize]{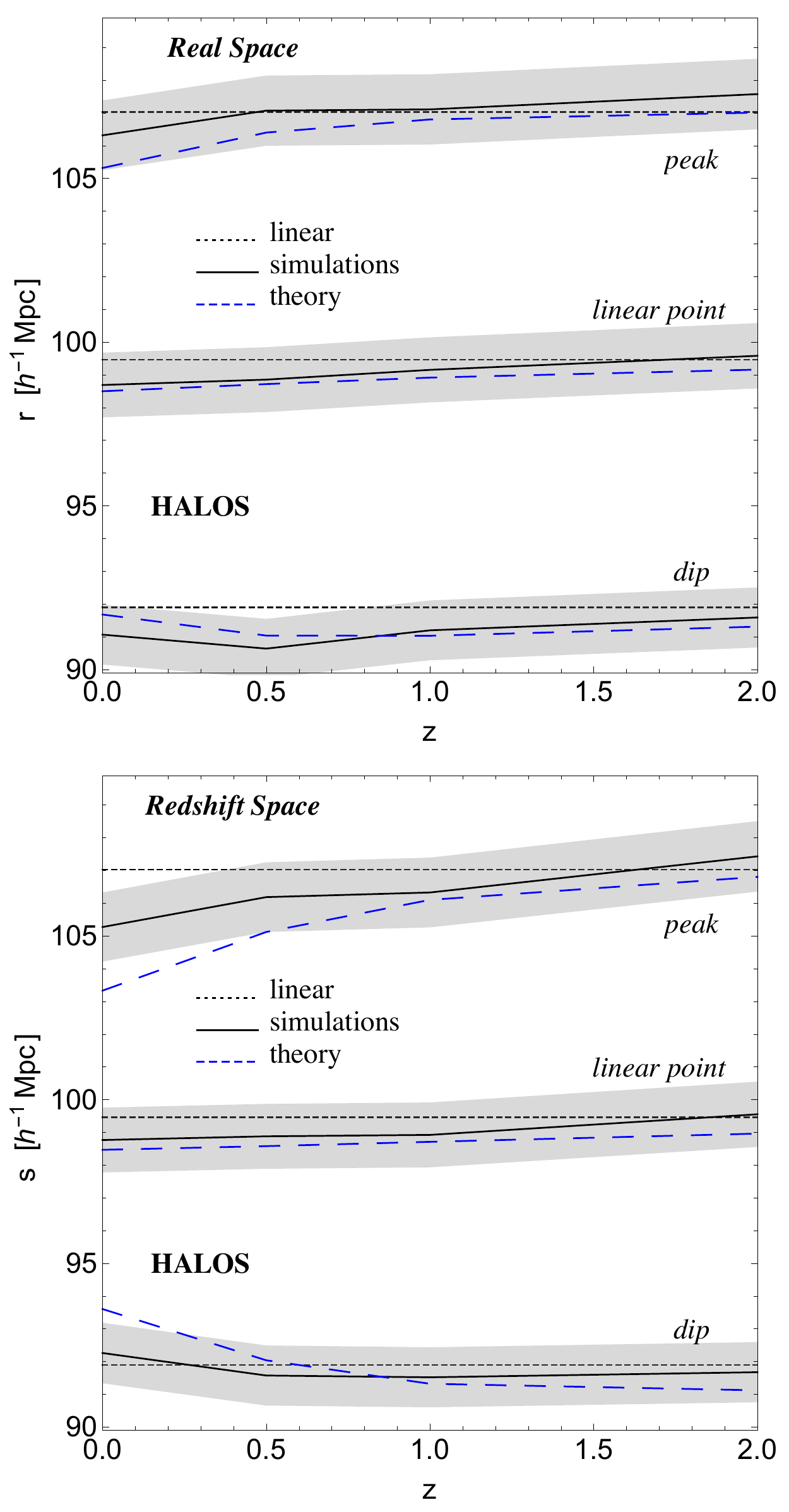}
\caption{\label{sim:theory:h} Same as Fig.~\ref{sim:theory:m} but for biased tracers.}
\end{figure}

Fig.~\ref{fig:rescaled:h} shows a similar analysis of the clustering of halos more massive than $M_{h}>1.37\times 10^{12} h^{-1}M_{\sun}$ at all the redshifts.  \footnote{\cite{2011PhRvD..84d3501S} identified the halos in each output using a friends-of-friends (FOF) group finder with linking length set to 0.2 times the mean particle separation; the mass limit above corresponds to halos with more than 20 particles.}  Symbols with error bars show the N-body results and smooth curves show eqns.~(\ref{peak:nl}) and (\ref{peak:nl:s}). For these, the linear bias $b^{2}_{10}(z)$ is fit to the simulation, and we found $b^{2}_{01}(z)$ to be negligible given the large errorbars. The top panel shows the real-space correlation function at four different redshifts, rescaled by the scale-independent linear growth factor as well as the linear bias $b^{2}_{10}(z)$. The linear bias values at $z=(0,0.5,1,2)$ are $b_{10}\simeq (1,1.29,1.72,2.95)$.  The bottom panel, shows the rescaled redshift-space monopole, where $\beta=f/b$. The nonlinear theory predictions are in reasonable agreement with the measurements.  Again, the error bars appear to be highly correlated, and, while the amplitude of $\xi$ is often discrepant by  up to 10\%, the locations of the standard rulers are much more robust. Fig.~\ref{sim:theory:h}, the analog of Fig.~\ref{sim:theory:m} shows a more quantitative comparison:  Notice that, especially in redshift space, the peak and dip scales differ more significantly from the measurements.  Nevertheless, the LP scale still agrees to better than $0.5\%$.  We conclude that the LP is indeed a good ruler even for biased tracers at late times in redshift space.  

Before we exploit this fact, it is worth making one final point.  Recall that the amplitude of $\xi^{lin}(s_{LP})$ is within about 3\% of $[\xi^{lin}(s_{d})+\xi^{lin}(s_{p})]/2$.  Figures~\ref{fig:rescaled} and~\ref{fig:rescaled:h} show that nonlinear evolution makes $\xi^{nl}$ less steep than $\xi^{lin}$, both in real and redshift space.  Hence, $\xi^{nl}(s_{LP})$ corresponds to $[\xi^{nl}(s_{d})+\xi^{nl}(s_{p})]/2$ at percent level precision, a factor of two improvement with respect to linear theory.

\section{Two consequences}\label{sec:use}

\subsection{Towards more accurate distance measurements}
We have made the point that the LP is mainly sensitive to the shift arising from the $r'/r$ factor (see Section \ref{subsec:m}).  We have found that a simple redshift independent $0.5\%$ correction removes this effect:  
\be
	s_{LP} = \frac{s_{p}+s_{d}}{2}\times 1.005 \, 
	\label{linear:lp}
\ee
and restores agreement with the linear prediction to $0.5\%$ at all redshifts both in real and redshift space. 

This is not the only option.  Since the peak and the dip scales are geometrical in linear theory, any linear combination of $s_{d}$ and $s_{p}$ which cancels the nonlinear shifts will return a geometric LP as well.  We have found that the optimum weighted sum at $z=0$ is 
\be
	s_{lin, w} = 0.7 s_{d} + 0.3 s_{p} ;
	\label{linear:w}
\ee
this combination returns the same linear theory value at all higher redshifts to $0.7\%$ (and can be corrected to $0.35\%$ using the same approach as in equation (\ref{linear:lp}).

One could even go further and employ a redshift weighted sum 
\be
	s_{lin}(z) = c(z) s_{d} +  d(z) s_{p} \, ,
	\label{linear:wz}
\ee
where $c(z)$ and $d(z)$ could be extracted from simulations so that $s_{lin}(z)$ would correspond to the linear theory prediction, i.e.~$1/2\lesssim c(z)\lesssim 7/10$ and $3/10\lesssim d(z)\lesssim 1/2$.  But we have not pursued this further as we feel we have made our main point:  using information from both the peak and the dip is potentially very powerful. Indeed not only do eqns. (\ref{linear:lp})--(\ref{linear:wz}) provide very stable scales for distance measurements, but comparing them at different redshifts provides useful cross checks.

\subsection{Using the LP to measure growth} 
So far we have made the point that the LP is a good standard rod.  However, its amplitude is also rather well-understood, so we now explore its use in measuring the linear theory growth factor.   

\begin{figure}
\centering
\includegraphics[width=1.0\hsize]{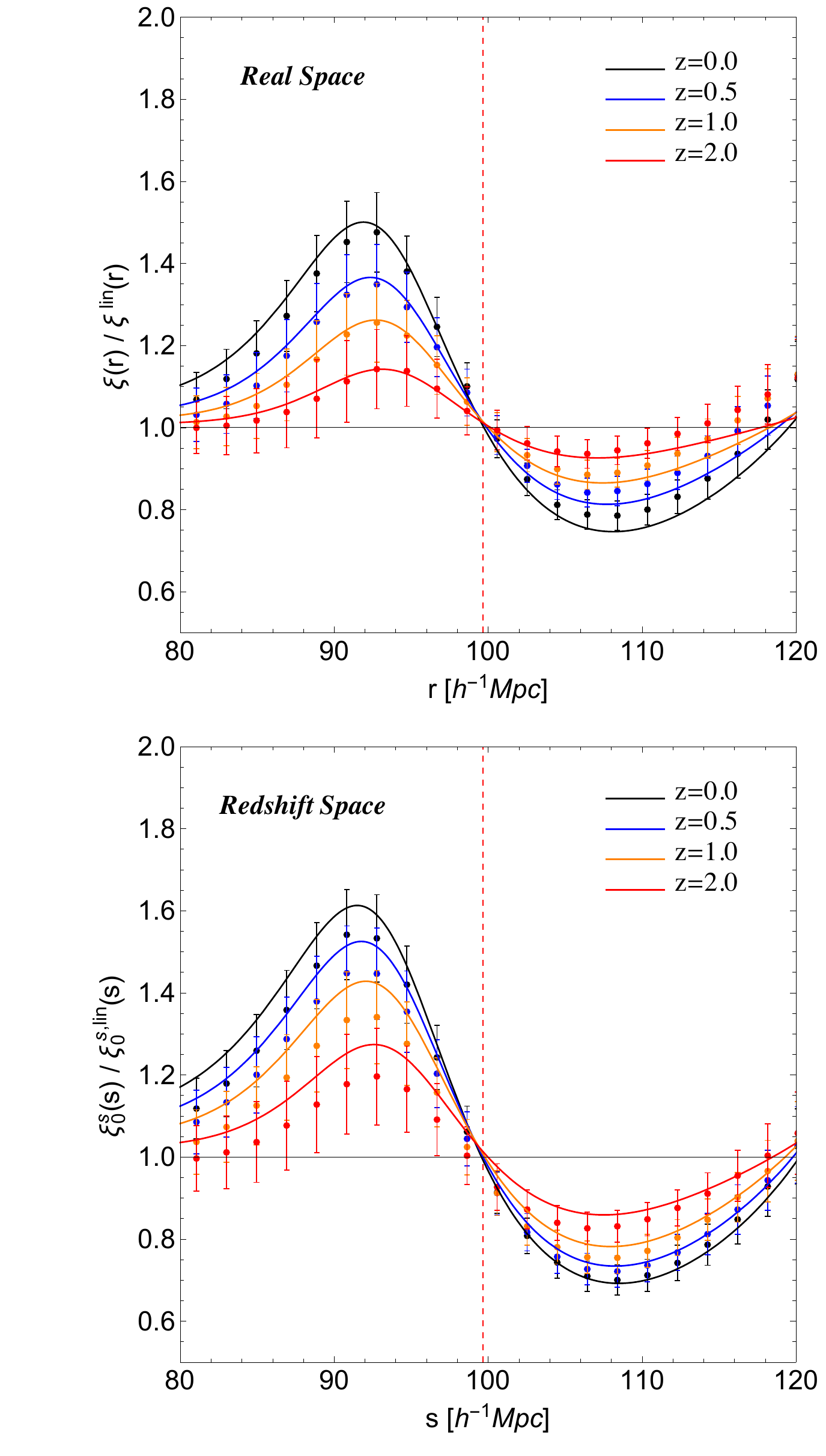}
\caption{\label{fig:relative} Ratio of the nonlinear correlation function to that in linear theory in real (top) and the redshift (bottom) space at four different redshifts. Smooth curves show the analytic model, symbols with error bars show the measurements and the vertical dashed red line shows the scale of the LP in linear theory.  At the LP, this ratio is very close to unity at all times.}
\end{figure}

\begin{figure}
\centering
\includegraphics[width=1.0\hsize]{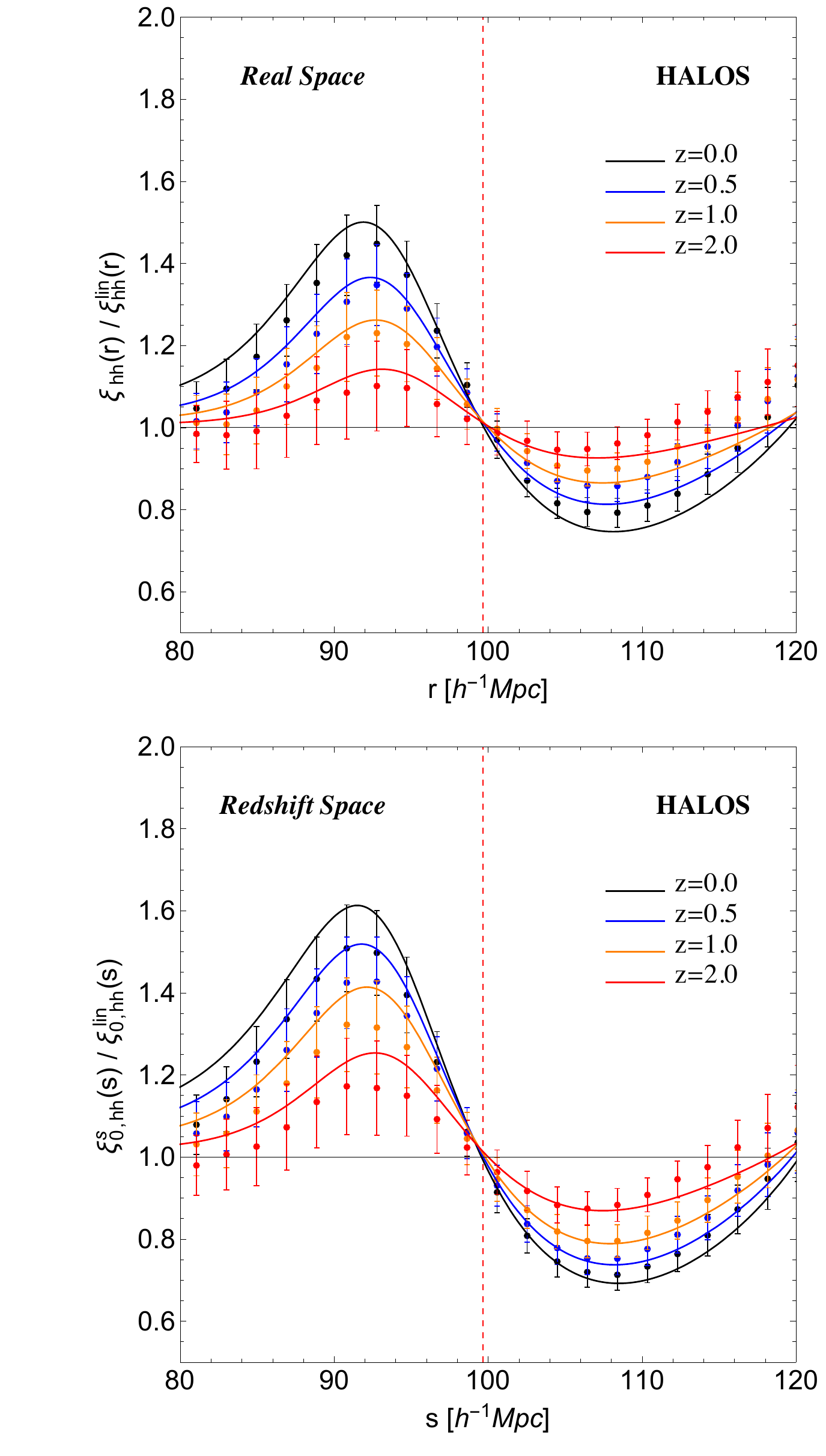}
\caption{\label{fig:relative:h} Same as Figure \ref{fig:relative}, but for biased tracers (halos more massive than $M_{h}>1.37\times 10^{12} h^{-1}M_{\sun}$).  }
\end{figure}

Over the BAO range of scales, our model predicts that the peak and dip heights should differ more from their linear theory values than do the peak and dip scales (c.f. Figures~\ref{fig:rescaled} and~\ref{fig:rescaled:h}).  However, at the LP, the amplitude of the correlation function is rather well described by linear theory.  To make this point, Fig.~\ref{fig:relative} shows the ratio of the nonlinear to the linear CF in the simulations (symbols with error bars) and in our model (smooth curves).  Although our model is able to describe the scale dependence of this ratio rather well (changes from unity can be as large as 40\% at $z=0$), it is remarkable how little the amplitude of the LP changes with redshift. Fig.~\ref{fig:relative:h} shows a similar analysis of biased tracers.  Again, at the LP, this ratio is within a few percent of unity.\footnote{As we anticipated in footnote 4 at the end of Section~4.2, there appears to be a scale larger than the BAO peak scale on which this ratio is close to unity at all redshifts even for biased tracers.  While it is not as close to unity as is the ratio on scale $s_{LP}$, this scale and the amplitude of $\xi$ on it potentially allow additional consistency checks of our $s_{LP}$ methodology which we leave for future work.}

For unbiased tracers, our analysis of the LP scale motivates the following estimator for the ratio of the linear growth factor at redshifts $z$ and $z'$:
\be
	\frac{D^{2}(z)}{D^{2}(z')} = 
	\frac{\xi^{lin}(s_{LP},z)}{\xi^{lin}(s_{LP},z')} 
	\backsimeq \frac{\hat{\xi}(\hat{s}_{LP},z)}{\hat{\xi}(\hat{s}'_{LP},z')}  \, .
	\label{growth:lp}
\ee
Here
 $\hat{s}_{LP}\equiv(\hat{s_{p}}+\hat{s_{d}})/2$, 
where carated quantities are the estimates inferred from observations.

While this is conceptually straightforward, it assumes errors on the measured amplitude are negligible.  Since this is rarely the case, one could reduce the effect of measurement errors by averaging the measured amplitude over more scales.  E.g., one might set  
\be
	\frac{D^{2}(z)}{D^{2}(z')} = \frac{\xi^{lin}(s_{p},z)+\xi^{lin}(s_{d},z)}{\xi^{lin}(s_{p},z')+\xi^{lin}(s_{d},z')} \backsimeq \frac{\hat{\xi}(\hat{s}_{p},z)+\hat{\xi}(\hat{s}_{d},z)}{\hat{\xi}(\hat{s}'_{p},z')+\hat{\xi}(\hat{s}'_{d},z')} \, ,
	\label{growth:sub}
\ee
or even 
\be
	\frac{D^{2}(z)}{D^{2}(z')} = \frac{\int_{s_{d}}^{s_{p}} \di x \, \xi^{lin}(x,z)}{\int_{s_{d}}^{s_{p}} \di x \, \xi^{lin}(x,z')} \backsimeq \frac{\sum\limits_{\hat{s}_{d}\leq x_{i}\leq \hat{s}_{p}} \hat{\xi}({x_{i},z)}/ N(z)}{\sum\limits_{\hat{s}'_{d}\leq x_{i}\leq \hat{s}'_{p}} \hat{\xi}({x_{i},z')}/ N(z')} \, ,
	\label{growth:sum}
\ee
where $N$ is the number of measurements done within the peak and the dip. 

In view of our discussion of the similarity of nonlinearities in real and redshift space, one could build similar estimators of the ratio 
$$
\frac{D^{2}(z)}{D^{2}(z')}  \frac{1+\frac{2}{3}f(z)+\frac{1}{5}f^{2}(z)}{1+\frac{2}{3}f(z')+\frac{1}{5}f^{2}(z')}
$$
from measurements of the redshift-space monopole.

Extending this to biased tracers is similarly straightforward.  We have already argued that the scale dependent bias terms matter little, so inserting the redshift space monopole of biased tracers for $\hat{\xi}$ in equations (\ref{growth:lp})--(\ref{growth:sum}) furnishes estimates of 
\be
	\frac{b_{10}^{2}(z)D^{2}(z)}{b_{10}^{2}(z')D^{2}(z')} \frac{1+\frac{2}{3}\beta(z)+\frac{1}{5}\beta^{2}(z)}{1+\frac{2}{3}\beta(z')+\frac{1}{5}\beta^{2}(z')}\, .
	\label{bias:s}
\ee
(The corresponding sums over the real space correlation function yield estimates of this quantity with $\beta\to 0$.)  

In all three versions of our estimator, the shift of the whole correlation function due to the $r'/r$ term in the convolution (see above) is irrelevant, provided that we consistently compare the peak-dip middle point \footnote{
 We ignore the correction introduced in equation~(\ref{linear:lp}).}.  
Therefore, these three estimators could also be used as a consistency check for the antisymmetry of the correlation function.  In $\Lambda$CDM, we expect this antisymmetry to be slightly broken by the mode-coupling contribution to the correlation function, although exploring this is beyond the scope of this paper.

While implementing the three estimators with actual data is beyond the scope of this paper, we believe that the following example is encouraging.  We focus on the halo redshift space monopole at $z=0$ and $z'=0.5$, since these biased tracers at low redshift are where the differences from linear theory are expected to be largest.  We then simply compare the theoretical predictions of $\xi^{s,nl}_{0, hh}$ with measurements in simulations $\xi^{s,sim}_{0, hh}$ (obtained by fitting a polynomial to the measured $\hat{\xi}$) and with the linear theory value.  For the halos in the simulation, the linear theory value of the ratio defined in~(\ref{bias:s}) is 0.9347.  Our nonlinear model predicts 0.9359 for all three estimators of this ratio, whereas the measured values are (0.9303, 0.9338, 0.9321).  
The agreement between the estimators as well as the nonlinear model with the linear theory value is better than $1\%$.  Clearly, an accurate analysis with actual BAO data is in order.

\section{Discussion and conclusions}
\label{sec:concl}

Since the first detection of the Baryon Acoustic peak in the galaxy correlation function \citep{Eisenstein:2005su}, tremendous effort, both  theoretical and observational, has been devoted to reducing systematic uncertainties and improving the statistics.  

With this in mind, we have searched for a special point in the correlation function whose scale and amplitude are least affected by nonlinear evolution.  We have found that the point which lies midway between the peak and the dip of the correlation function on BAO scales, which we call the linear point $s_{LP}$ (see, e.g., Figure~\ref{fig:BAOfeature}), is shifted in scale from its linear theory value by less than $0.5\%$.  Previous work has shown that nonlinear evolution smears out the height and shifts the scale of the peak.  We showed that there are similar effects on the dip, but that these nonlinear corrections cancel out almost exactly at the linear point (Figure~\ref{fig:shift}).  We provided analytic arguments for why this same LP is special for unbiased as well as biased tracers, in real and redshift space (Section~4), and showed that this was indeed the case in simulations (Figures~\ref{fig:rescaled} and~\ref{fig:rescaled:h}).  

We believe that this motivates use of the LP for distance measurements, as it may remove the need to fit broadened correlation function templates to the measurement in order to infer the acoustic scale as is currently done \citep[e.g.,][]{2009MNRAS.400.1643S, Sanchez:2012sg}.  For example, estimating $s_{LP}$ from the published BOSS DR11 galaxy correlation \citep[Figure~10 of][]{2014MNRAS.441...24A}, following the same simple procedures we used to estimate it in the simulations presented in the main text (fitting a high-order polynomial to the measured $\xi$) and applying equation~(\ref{linear:lp}) yields $95.1h^{-1}$~Mpc. For comparison, $s_{LP}$ in the fiducial model of \cite{2014MNRAS.441...24A}, divided by $\alpha=1.03$ as their Table~7 suggests, also yields $95.1h^{-1}$~Mpc.  We leave a more careful implementation of this procedure (e.g., accounting for covariances between bins, reconstruction, etc.) to future work.

A nontrivial consequence of the cancellation of nonlinear effects is that it is not just the scale of the LP which remains fixed:  its height is also given by linear theory, to percent level precision, even at late times, for unbiased and biased tracers in real and redshift space (Figures~\ref{fig:relative} and~\ref{fig:relative:h}).  We exploit this interesting property to motivate three estimators of the linear theory growth -- equations~(\ref{growth:lp})--(\ref{growth:sum}) -- and to compare our findings with data. 

In practice, our simplest estimators of the growth constrain the combination $b_{10}\,D$ rather than $b_{10}$ and $D$ separately.  We expect the width of the `nonlinear' smoothing to be approximately independent of $b_{10}$, so it may be that combining our LP-derived constraint on $b_{10}\,D$ with the more traditional estimators of the acoustic scale and its broadening will allow one to infer $b_{10}$ and $D$ separately.  Breaking this degeneracy is the subject of work in progress.  

Our detailed results for the CF structure in the BAO regime could be modified by non-standard physics. At the linear level we expect that the presence of massive neutrinos, some form of dark energy, or some modification of gravity will change the discussion of Section \ref{sec:lin} \citep{Thepsuriya:2014zda}. On the other hand, when we deal with nonlinearities we should be careful to make some distinctions.  The BAO dip-peak region in the presence of massive neutrinos is well described by the same physics as $\Lambda$CDM \citep{2015arXiv150507477P}.  Quintessence and clustering quintessence also show the same nonlinear propagator as the standard case \citep{Anselmi:2014nya}, the transition from large to small scales being smooth. 
On the other hand, a preferred scale often appears in modified gravity models, so equation~(\ref{xiZel}) can no longer be  applied naively \citep{Scoccimarro:2009eu}.  It is not obvious that the antisymmetric features of the correlation function will be preserved in general. In this regard, precise measurements by future surveys of departures from the antisymmetry in the BAO range of scales would be a clear signature of non-standard gravity.

\section*{Acknowledgments}
We thank M. Sato and T. Matsubara for providing the clustering measurements from their simulations. SA and GDS are supported by a Department of Energy grant DE-SC0009946 to the particle astrophysics theory group at CWRU.


\bibliography{MyBib}{}

\label{lastpage}

\end{document}